\documentclass[a4paper, oneside, twocolumn, notitlepage, 10pt]{extarticle_ecoc}
\usepackage{ecoc}
\usepackage{caption}

\begin{document}
\selectlanguage{american}    


\title{Field-Trial of Machine Learning-Assisted Quantum Key Distribution (QKD) Networking with SDN}%


\author{Y. Ou, E. Hugues-Salas, F. Ntavou, R. Wang, Y. Bi, SY. Yan, G. Kanellos, R. Nejabati, D. Simeonidou}

\maketitle                  


\begin{strip}
 \begin{author_descr}

     \textsuperscript{} High Performance Networks group, School of \& Computer Science, Electrical and Electronic Engineering, and Engineering Maths, University of Bristol, BS8, 1UB, UK,
   \textcolor{blue}{\uline{yanni.ou@bristol.ac.uk}} 



 \end{author_descr}
\end{strip}

\setstretch{1.1}


\begin{strip}
  \begin{ecoc_abstract}
We demonstrated, for the first time, a machine-learning method to assist the coexistence between quantum and classical communication channels. Software-defined networking was used to successfully enable the key generation and transmission over a city and campus network.
  \end{ecoc_abstract}
\end{strip}

\section{Introduction}
\setlength{\parskip}{-0.2em}
Integrating quantum key distribution (QKD) with existing ``conventional'' DWDM fibre networks is crucial for bringing the implementation and commercialisation of quantum networks into reality \cite{ref1}. In addressing the coexistence of quantum-encoded photons with high-intensity classical signals in C-band, the main challenge is the photon-induced noise falling in the quantum channel (Ch-QKD) that deteriorates the Ch-QKD quality and the generated secret key rate (SKR). Main noise sources include: i) Raman Scattering (RS), ii) the non-filtered photons from classical channels (Ch-C), iii) the ``in-band'' noise photons induced by nonlinearity, e.g., FWM, and iv) the spontaneous emission (ASE) photons from optical amplifiers. Recent C-band quantum coexistence demonstrations \cite{ref2} report on point-to-point transmission for given wavelength allocations. However, software-defined networking (SDN)-enabled network-wide transmission of co-existing Ch-QKDs and Ch-Cs would allow for further noise suppression by dynamically selecting noise optimised network paths but requires to consider the dynamic allocation of wavelengths in spectra, being typical in current deployed optical networks. Traditional approaches \cite{ref3} to estimate the Ch-QKD in-band noise are difficult to implement in a network with realistic spectrum utilisation, considering a larger number of Ch-Cs located at various wavelengths, in various channels spacing, modulated by hybrid formats and carrying varying data rate.

Therefore, in this paper, we apply for the first time a supervised machine learning (ML) method to estimate the Ch-QKD performance (noise, SKR and Quantum Bit Error Rate - QBER) in the presence of Ch-Cs in various quantities, spectrum allocations, launch power and channels spacing in C-band. Training sets for ML were collected from 3 aspects: one from a laboratory testbed, and two from field-trial optical networks: the campus and the Bristol city networks (CityNet). Several ML models were compared to evaluate prediction outcomes. Moreover, based on ML, we provide a scheme for predicting optimum optical channel allocation, enabling the SDN platform to spectrally re-allocate the channels for constant performance. To verify the SDN-enabled ML prediction, field-trial networks were adopted with robust performance of Ch-QKD, showing the feasibility of the ML application.

\begin{figure*}[h!]
\begin{tabular}{cc}
 \hspace*{-10pt}
 \includegraphics[width=0.55\linewidth]{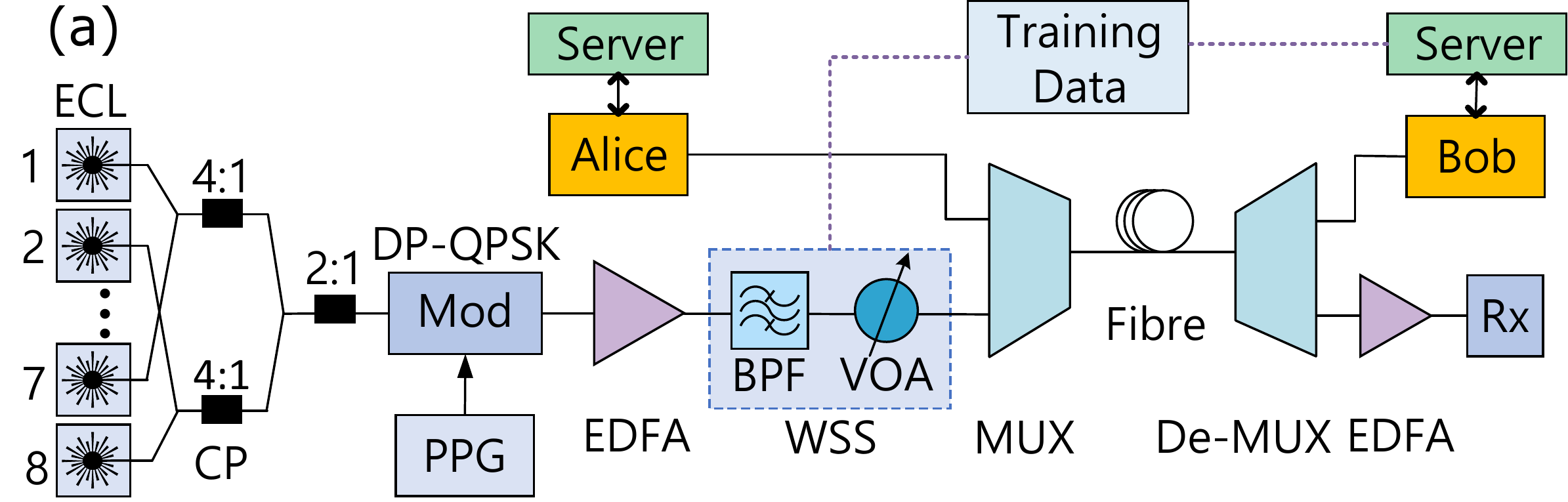}&
 \hspace*{-16pt}
 \includegraphics[width=0.46\linewidth]{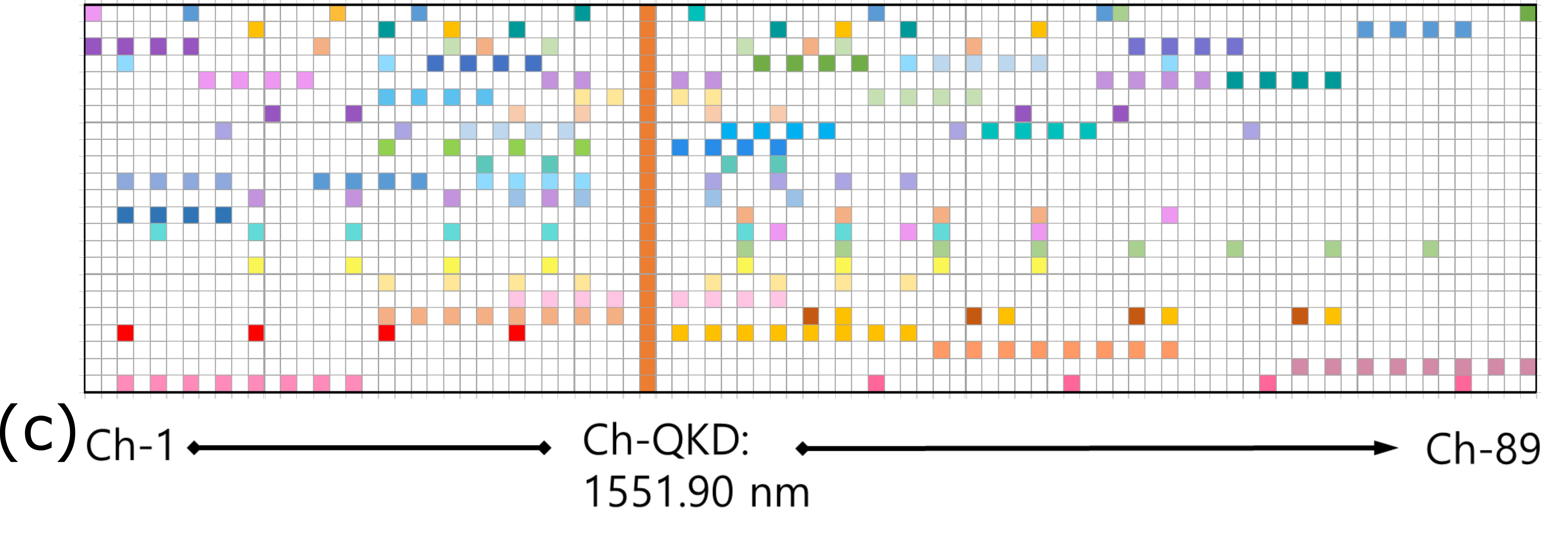}\\
 \vspace*{-15pt}
\end{tabular}

\begin{tabular}{cc}
 \hspace*{-10pt}
 \includegraphics[width=0.43\linewidth]{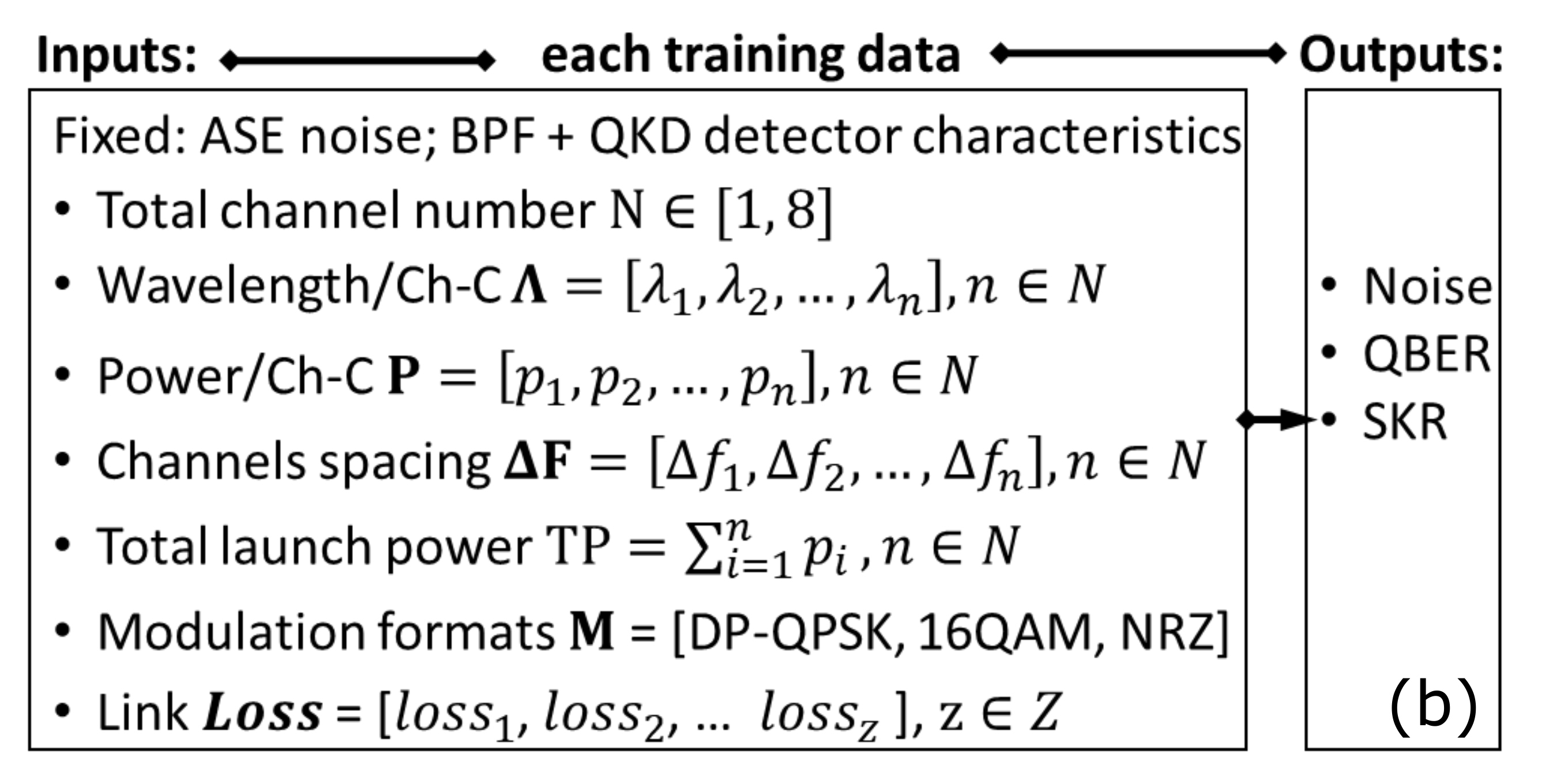}&
 \hspace*{-13pt}
 \includegraphics[width=0.57\linewidth]{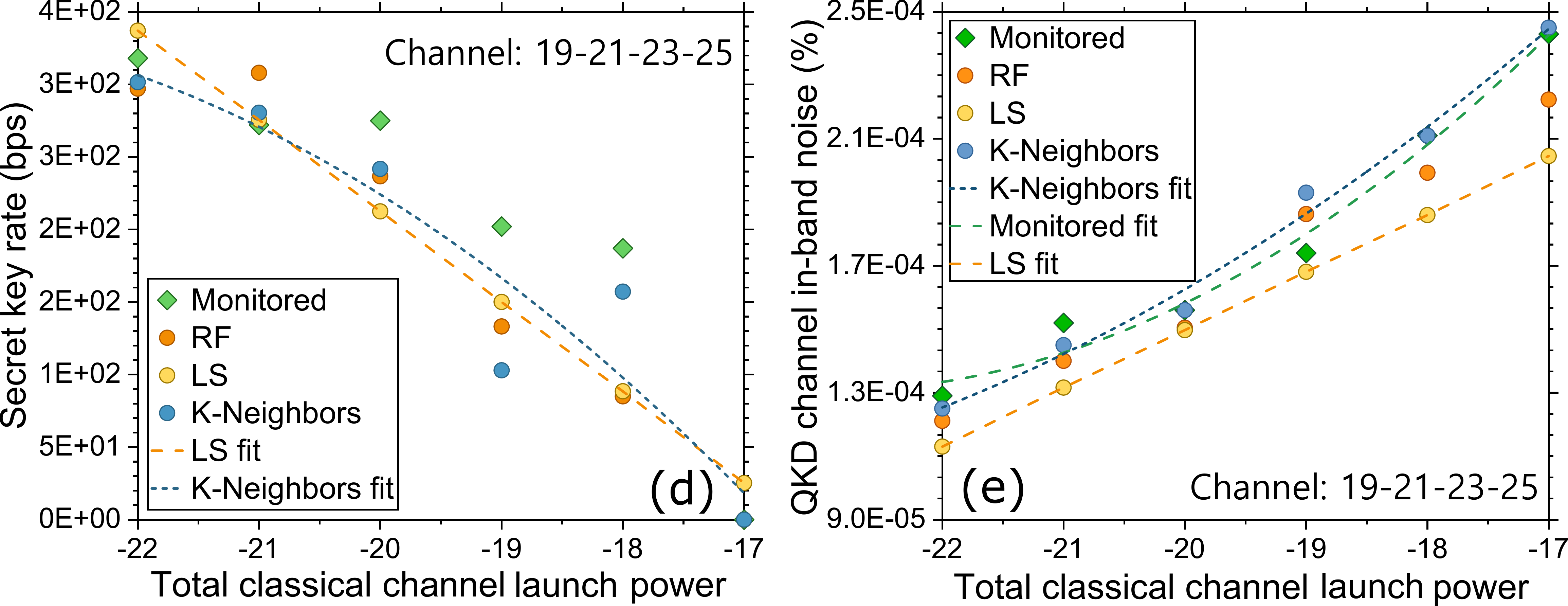}\\
\end{tabular}

\captionsetup{justification=normal}
\caption{a) Testbed schematic for collecting training sets over 3 fibre types: lab, campus and CityNet; b) in-/outputs composition of training sets for supervised learning; c) C-band distribution of Ch-Cs in different $\Delta \bold{F}$ and $n=(1,4,8)$ for composing training sets, fixed Ch-QKD at 1551.9nm, each colour per line is a $c \in \bold{C}$ in the ITU-T DWDM C-band; d) predicted SKR and e) in-band noise, for two given channel conditions using ML models.}
\end{figure*}

\section{Training data collection and ML models}

Predicting the Ch-QKD performance is a regression problem and we use supervised learning to attack. Therefore, each training instance in the sets is formed by the input(s) and their corresponding output(s). The outputs are the parameters that reflect the Ch-QKD performance to be predicted after learning. The inputs describe the attributes of Ch-Cs and are concluded by Fig.1b. $N$-channels, central wavelength(s) of $n \in N$ channel(s) and channels spacing are represented as vectors $\bold{\Lambda}$ and $\bold{\Delta}$. The per Ch-C launch power and the total power for $n$-channels are also input features shown as $\bold{P}$ and $TP \in [-15, -26]$ dBm, determining the noise induced by RS. Combinations of these features $\bold{C}=(n,\bold{\Lambda},\bold{\Delta},\bold{P})$ will induce different degrees of Four-wave mixing (FWM) effects. A selection of combinations are distributed in Fig.1c to form the training sets, emulating different spectrum utilisation in an optical path. Other fixed input attributes are: the ASE noise accompanied by Ch-Cs, emulating the channels adding in a network; the modulation formats and filtering are attributes relevant to the power and polarisation of Ch-Cs; and the QKD detector's characteristics define the Bob's efficiency. For the outputs, Bob's detection probability (dark count) is used to evaluate the Ch-QKD. 

The testbed in Fig.1a is implemented to collect the training sets, where $N (=8)$ Ch-Cs are emulated by tunable lasers and the 32GBd DP-QPSK modulator. Since the Ch-QKD generated from IDQ Clavis2 is fixed at 1551.90nm  corresponding to the ITU-T Ch-35, we select the wavelengths $\Lambda$ whose mixing products are more likely to fall in the Ch-QKD according to $f_{ijk}$ and $f_{low(high)} = 2f_{min(max)}-f_{max(min)}$ to obtain the most FWM impact induced by $1/2(N^3-N^2)$ number of Ch-Cs. The channels will be filtered and amplified once before going to the WSS for the attenuation and filtering to vary their launch power individually. Three fiber types are used in this testbed: a 1-km SSMF spool, a 1-km campus network and a 8.6 km CityNet, with corresponding end-to-end losses of 9.5, 10.2 and 9dB. After transmission, they were demultiplexed and the Ch-QKD is bandpass filtered for Bob's detection of noise, SKR and QBER. We obtained 5 training sets (each has 164 instances) and 5 validation sets (each has 43 instances). The validation sets are excluded from the training sets to evaluate the prediction of Ch-QKD under the unforeseen conditions. Considering our obtained training sets are less abundant due to the limited lab set ups, the following supervised ML models are selected for this work to compare, including Random Forest Regression (RF), Least Squares (LS), K-neighbours Regression (KN), Lasso and Ridge Regression. RF is used to set a benchmark, as it was proved to be the best family of classifiers among the 179 classifiers from 17 families \cite{ref4}. 

\begin{figure*}[h]
   \centering
        \includegraphics[width=1\textwidth]{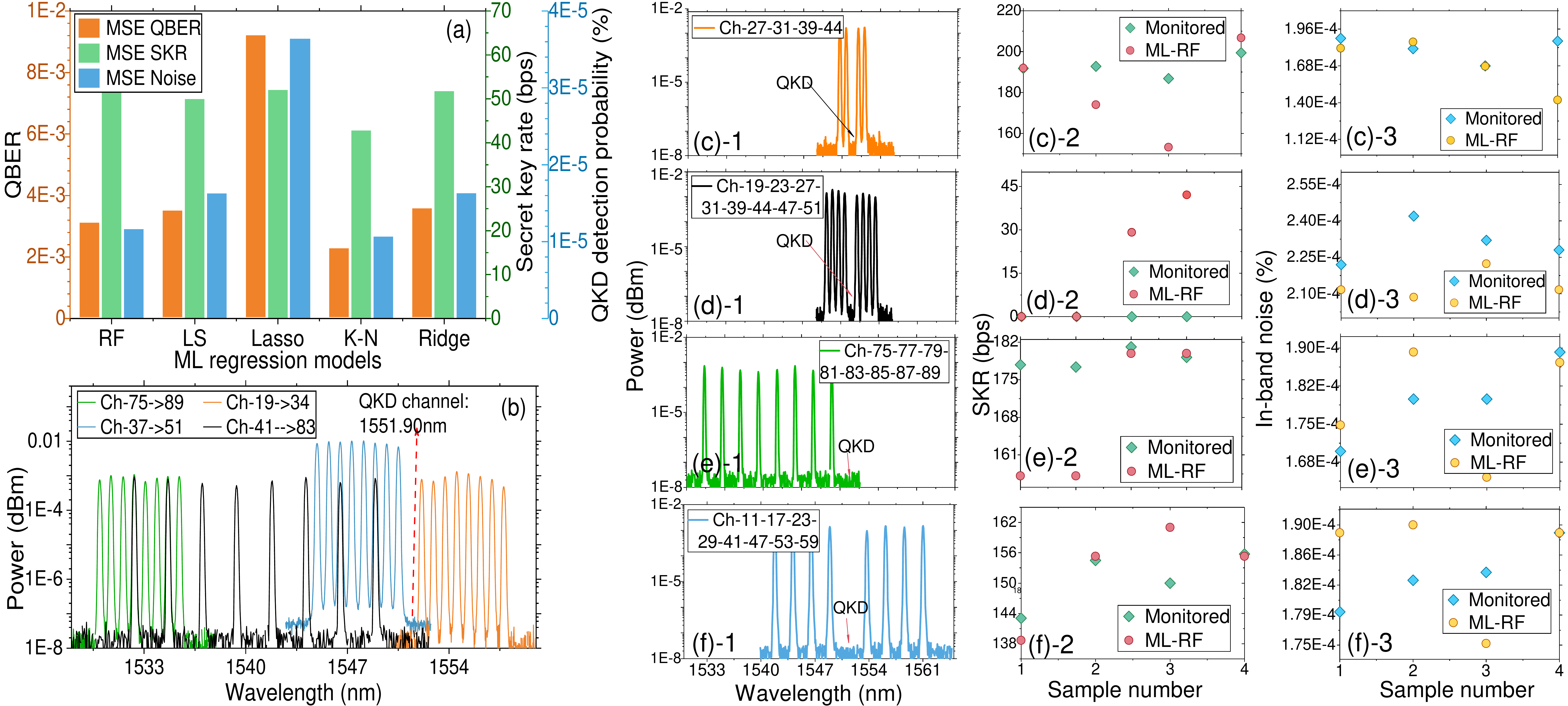}
    
    \captionsetup{justification=normal}
    \caption{a) Comparison of MSE prediction of SKR, QBER and noise from RF, LS, Lasso, KN and Ridge ML models; b) training sets allocation for 8-channels with Ch-QKD in 1551.9nm; c)-f) monitored and predicted SKR and noise from the use case in the c)1-3: initial stage, d)1-3: stage-1, e)1-3: stage-2, and f)1-3: stage-3, with the adding of new Ch-Cs and their (re-)allocation.}
    \label{fig:figure1}
\end{figure*}

Fig.2a compares the results from 5 ML models in predicting noise, QBER and SKR for validation sets. The mean square error (MSE) between the predicted and the monitored values were calculated. K-Neighbours shows the best overall performance but mildly transcends RF. Fig.2b reveals the examples of spectrum allocations for different combinations $c$ of Ch-Cs. Fig.1d-e further elaborate the ML prediction based on two input examples from validation sets. The results show that the RF's prediction is in high variance, while the averaged errors are small as it is a model ensemble method. KN performs better than RF while suffers from the ``curse of dimensionality'' as it is non-parametric. This is concealed with our 7-dimensional input attribute space, but its performance will drop significantly with increased dimensions. On the other hand, LS is easy to fit as it is under a linear assumption. Ridge, as a regularized linear regression, has a comparable performance to LS as it avoids the extreme values of the parameters to give a prior belief.

\section{SDN based network use case using ML for QKD and classical channels optimisation}

Fig.3a shows the use case testbed that realises the optical network (re-)configuration to guarantee both QKD and Ch-Cs transmission, by adding ML as an application with SDN. The optical plane is similar with Fig.1a, and the encoded photon from Alice is multiplexed into the network through an access node. In the control plane, a monitoring application is developed to continuously monitor the state of the above optical links in terms of channels allocation and optical power levels. These values are sent from the SDN-enabled SSS to the SDN Controller (OpenDaylight) using the extended OpenFlow (OF) and then handled by the application. A MongoDB database is designed to store monitoring values, allowing the ML application to extract them for predicting the Ch-QKD in-band noise, SKR and QBER. After the prediction, the configuration data is sent to the SDN controller, which will appropriately configure the switches through their OF Agents.
\begin{figure}[b!]
  \vspace*{-19pt}
  \includegraphics[width=0.49\textwidth]{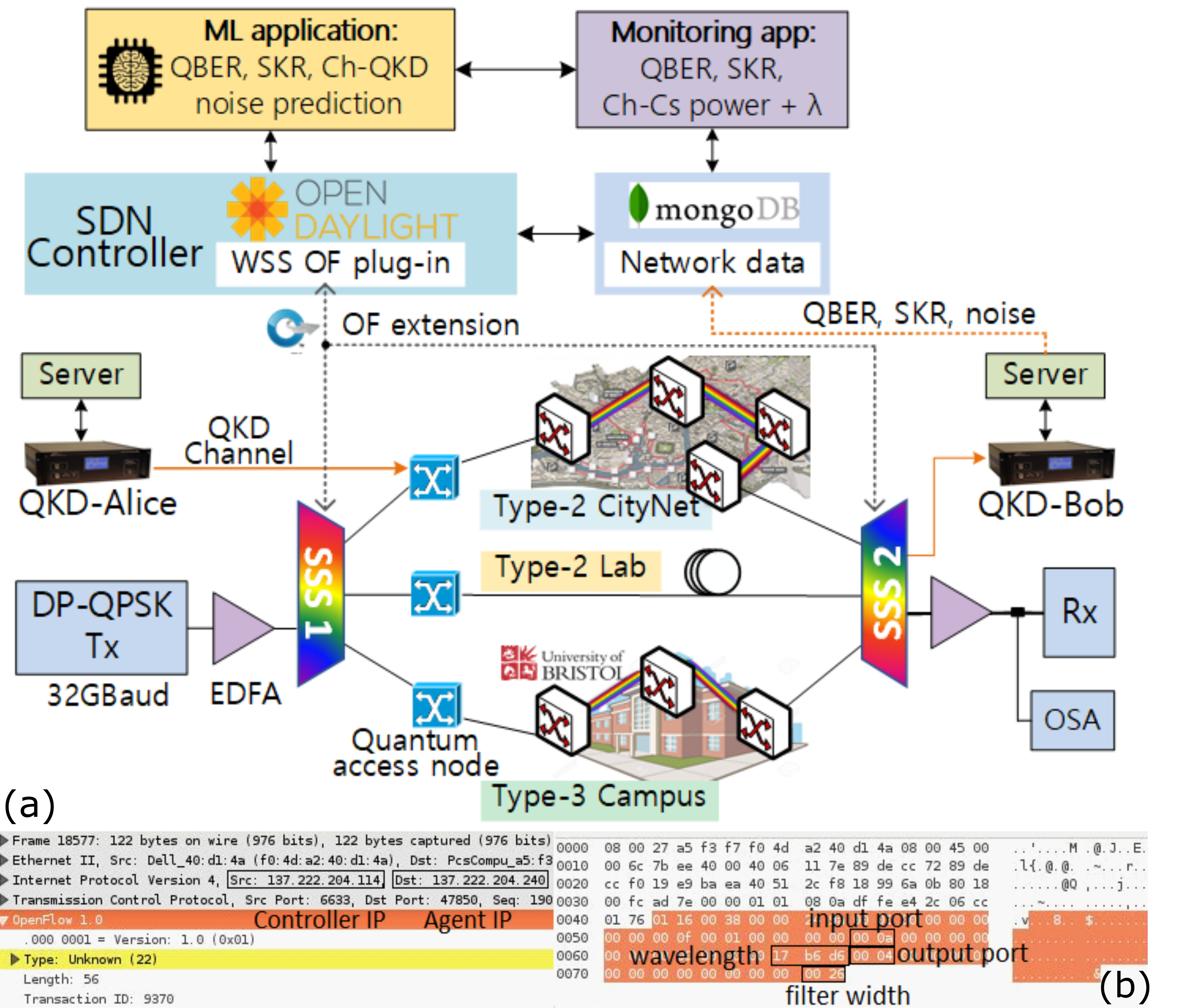}
  
  \captionsetup{justification=normal}
  \caption{a) Field-trial testbed for the SDN-enabled network use case with ML and network monitoring; b) extended OF message for SSSs to re-allocate Ch-Cs in different stages.}

\end{figure}

In this use case, we consult the trained ML algorithm first on the optimal path out of the three to transmit the Ch-QKD and four Ch-Cs (Fig.2c1-3 initial stage). Then, 4 more Ch-Cs are added in the link to emulate the network changes, which is observed by the monitoring application with the increase of power (Fig.2d1-3 stage-1). The ML is called to predict the noise, SKR and QBER based on the current channel condition. If the predicted values are worse than a requested threshold, two alternative options will be invoked: i) to re-allocate the wavelengths of the 8 Ch-Cs in the same path for reducing their impact on the Ch-QKD; ii) to switch to another path. In this case, option i) is adopted, and two re-allocation choices are decided: Ch75-77-70-81-83-85-87-89 and another back up channel Ch11-17-23-29-41-47-53-59. The latter fits in a more spectrum fragmented path. Fig.2e-f  reveals the noise and SKR prediction. The decision is then sent to the SDN controller, which executes the corresponding re-configurations of the optical devices within the network to re-allocate the Ch-Cs, including laser wavelengths and SSS. After re-allocation, Ch-QKD is validated by monitoring the noise and SKR in comparison to the predicted values shown in Fig.2e1-3 stage-2 and Fig.2f1-3 stage-3. Fig.3b indicates an OF message example sent to configure the SSS, depicting the in-/output ports A and 4 of the SSS, the new wavelength of 1554.134nm and the 38 GHz filter width.


\section{Conclusion}

We predict the Ch-QKD quality in a QKD-DWDM network with multiple ML models, showing that KN is the most accurate model. The efficiency of the ML prediction is demonstrated in a field-trial SDN-enabled optical network.

\section{Acknowledgements}
The work acknowledges EP/M013472/1: UKNQT, EP/L020009/1 TOUCAN and METRO-HAUL.

\bibliographystyle{abbrv}
\begin{spacing}{1.1}

\end{spacing}
\vspace{-4mm}

\end{document}